# A Novel Generative Artificial Intelligence Method for Interference Study on Multiplex Brightfield Immunohistochemistry Images

Satarupa Mukherjee, Jim Martin, Yao Nie

Roche Diagnostic Solution, Computational Science and Informatics, Pathology Lab Solutions, Santa Clara, CA, USA

**ABSTRACT**

Multiplex brightfield imaging of pathology tissue slides promises to enable a deeper understanding of the tumor microenvironment. However, quantitative interpretation of multiplexed images through visual assessment is time-consuming and error prone. To accurately analyze multiple biomarkers, identifying which chromogen is present on each cell is crucial, which requires advanced color unmixing methods to create synthetic singleplex images for each biomarker. This task is particularly challenging when more than two biomarkers are present in the same cellular compartment. In this project, two representative biomarker sets were selected as assay models - cMET-PDL1-EGFR and CD8-LAG3-PDL1, where all three biomarkers can co-localize on the cell membrane.

To address the above-mentioned challenges, we developed a generative artificial intelligence (AI) method for color unmixing on the triplex assay models. The approach was used to aid interference study, which was an essential step for optimizing the multiplex assays by detecting indications of chemical interference. In this study, chromogen analogues were used to model for potential gross interferences that could occur among the three chromogens that were used to develop the triplex assay. Specifically, for each assay model, we developed three cycle-generative adversarial network (cycle-GAN) models to map an input triplex image to each of the three stains: Tamra (purple), QM-Dabsyl (yellow) and Green, respectively. The novelty of our cycle-GAN network is that the input to the network are images in the optical density domain instead of conventionally used RGB images. This is shown to help reduce the blurriness of the synthetic singleplex images, which was often observed when the network was trained on RGB images.

The cycle-GAN models were validated on 10,800 lung, gastric and colon images for the cMET-PDL1-EGFR assay and 3600 colon images for the CD8-LAG3-PDL1 assay. Visual as well as quantified assessments demonstrated that the proposed method is effective and efficient when compared with the manual reviewing results and is readily applicable to various multiplex assays.

**Keywords:** cycle-GAN, Color unmixing, Digital Pathology, Artificial Intelligence, Multiplex brightfield imaging

## 1. INTRODUCTION

Multiplex brightfield imaging offers the significant advantage of simultaneously analyzing multiple biomarkers on a single tissue slide, as opposed to multiple different single biomarker labeling on multiple consecutive slides. This enables investigating interactions between biomarkers more precisely to gain a better understanding of the tumor microenvironment and thus improved diagnostic and prognostic capabilities.

To establish the ability to accurately analyze multiple biomarkers that are localized in the same cellular compartment, two representative biomarker sets were selected as assay models - cMET (tyrosine-protein kinase Met)-PDL1 (programmed death-ligand 1)-EGFR (Epidermal growth factor receptor) and CD8 (cluster of differentiation 8)-LAG3 (lymphocyte-activation gene 3)-PDL1. For analyzing a multiplexed immunohistochemistry (mIHC) assay, one of the most crucial steps is to identify the individual chromogens on each cell in the stained slides. This is a very challenging task for both the selected assay models, because co-localization of all three biomarkers in each model could occur on the cell membranes. Due to the fact that membrane stains often overlap with the nuclear counterstain, resulting in four different colors mixed at the same location, this problem cannot be solved by the conventional Lambert's Law based color deconvolution method [1] that can only unmix up-to three colors. Thus, an advanced color deconvolution (unmixing) method is demanded, so that we can create equivalent singleplex images for each biomarker for the downstream analysis. These singleplex images created from a triplex image are referred to as synthetic singleplex images.

In this project, we developed a generative Artificial Intelligence (AI) approach for color unmixing the slides stained by the abovementioned triplex assays to aid assay interference study. The purpose of the interference study was twofold. First, chromogen analogues were used to model for potential gross interferences that could be occurring between the three chromogens that were used to develop the triplex assay. Once gross interference evaluations were complete, order of biomarker addition was established based on least predicted interference using the gross interference model. These triplexes were then used for color unmixing. Once the unmixing models were finalized, these were used with the predicted low chemical interference configuration to establish any indication of chemical interference as compared to the adjacent single stained slides.

One widely used color unmixing method for unmixing more than three colors is non-negative matrix factorization [2], which has been investigated extensively in our project. This approach, however, heavily relies on stains' statistics within an image and may fail when multiple stains are co-localized. Additionally, these methods are generally difficult to optimize, often requiring in-depth prior knowledge, constraints and extensive parameter tuning. To overcome these disadvantages, we developed a cycle-Generative Adversarial Network (cycle-GAN) [3] method for unmixing the triplex images generated from the above-mentioned assays. Three different models were designed for each of the three stains – Tamra (purple), QM-Dabsyl (yellow) and Green. This method learned the stain separation function from the data and was less sensitive to variations in the stain level. Whereas conventional unmixing algorithms perform poorly when applied to different protocols, tissues, subjects, cancers or sites, this cycle-GAN based approach could robustly generate synthetic singleplex images across different contexts.

## 2. METHODOLOGY

The cycle-GAN [3] is a technique for training image-to-image translation models without requiring paired examples. In general, the model consists of two generators and discriminators, all of which enable learning of the mapping between the source and target domain by using the collection of images from each domain that may not be directly related. In our use case, a multiplex image and the associated singleplex images were considered as source and target images, respectively. Figure 1 illustrates an example of the cycle-GAN architecture. As shown in the figure, the approach trains one model to map an input triplex image to the singleplex image that corresponds to one biomarker, which results in three distinct models in total to perform the entire unmixing task. A notable novelty of our approach was that we employed images in the optical density (OD) domain instead of conventionally used RGB images as the input to the network. This brought a few potential advantages for the model training due to the inherent properties of the OD domain:
- The OD transfer enhances the contrast between different light absorbing materials, making subtle differences more apparent. This improved contrast can help deep learning models better distinguish between different components, leading to more effective feature extraction.
- The OD domain allows for linear decomposition of mixed signals. For deep learning models, this can mean that the underlying patterns and structures are more easily captured and represented, potentially helping the deep learning model to learn more effectively and efficiently.

- The OD domain aligns well with the physical principles of light absorption by biological stains. Training models in this domain can leverage this domain knowledge, leading to more biologically and physically meaningful feature learning.

The use of OD domain helped reduce the blurriness of the synthetic singleplex images which was often observed when the network was trained on RGB images.

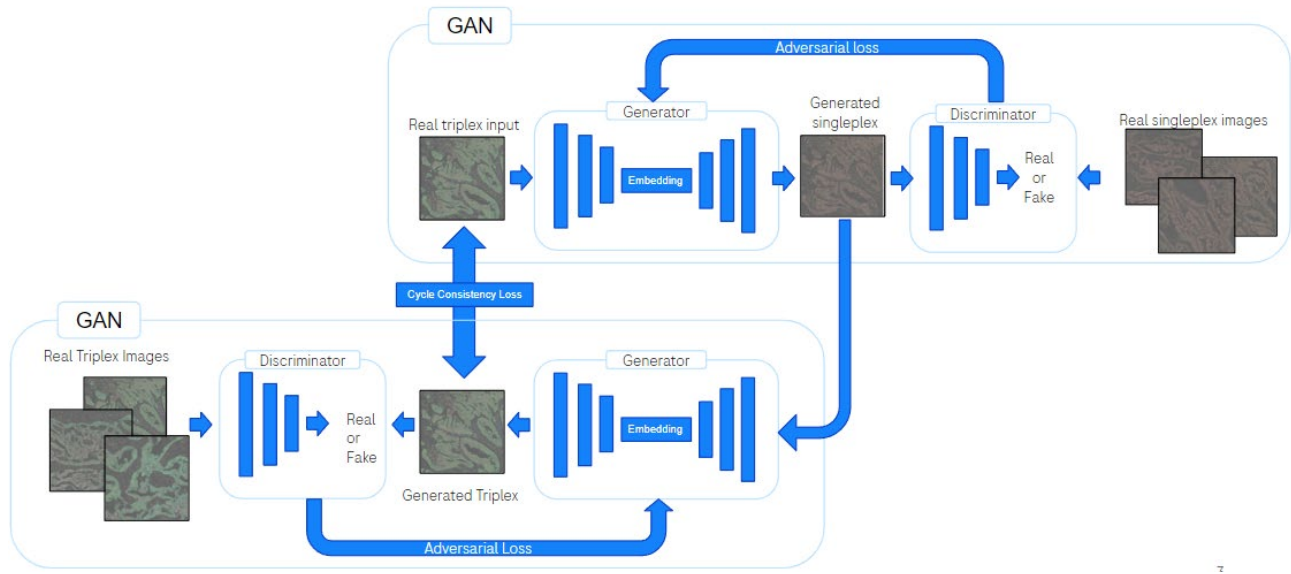

**Figure 1. Overview of the cycle-GAN model architecture for generating synthetic singleplex images for one stain**

The cycle-GAN was trained and evaluated on two different assay models: cMET-PDL1-EGFR and CD8-LAG3-PDL1, respectively. For each model assay, three different models were trained for the three stains, i.e.,Tamra, QM-Dabsyl and Green, respectively.

For the cMET-PDL1-EGFR assay, a total of 12 different whole slide tissue images scanned at 0.5micron/pixel were used, 4 each from lung, gastric and colon, respectively. Slides from each tissue sample were stained with either a single biomarker (eg. referred to as single stain) or hematoxylin. The single stained/hematoxylin slides were used as proxy-ground truth for evaluation purposes and also as input to the discriminator.

On each slide, 10 fields-of-view (FOV; image size of 1586x1540 pixels) were selected by a pathologist. From each FOV, either triplex or singleplex, 30 patches of image of size 256x256 were randomly cropped and were split into 80-10-10 ratio for training, testing, and validating the models respectively.

For the CD8-LAG3-PDL1 assay, a total of 4 different slides were used from colon tissue samples. Similarly, as the previous dataset, on each slide, 10 FOVs (image size of 1586x1540) had been selected by a pathologist. From each FOV, 30 patches of image size of 256x256 had been randomly cropped and were split into 80-10-10 ratio for training, testing, and validating the models respectively.

3. **RESULTS**

The cycle-GAN models were validated on 10,800 lung, gastric and colon images of cMET-PDL1-EGFR and 3600 colon images of CD8-LAG3-PDL1 having various intensities of Tamra, QM-Dabsyl and Green. In the following sections, visual and quantitative results as well as pathologist evaluations are presented for each of the assay to demonstrate the correlations between the synthetic singleplexes (created by the cycle-GAN method) and the adjacent singleplexes for each of the stains.

## 3.1. RESULTS ON CMET-PDL1-EGFR

Figures 2-4 show visual results of the method for all the three stains of the cMET-PDL1-EGFR assay. The leftmost image in each of the figures shows the triplex, the middle image shows the generated synthetic singleplex obtained from the cycle-GAN and the rightmost image shows the corresponding adjacent singleplex image.

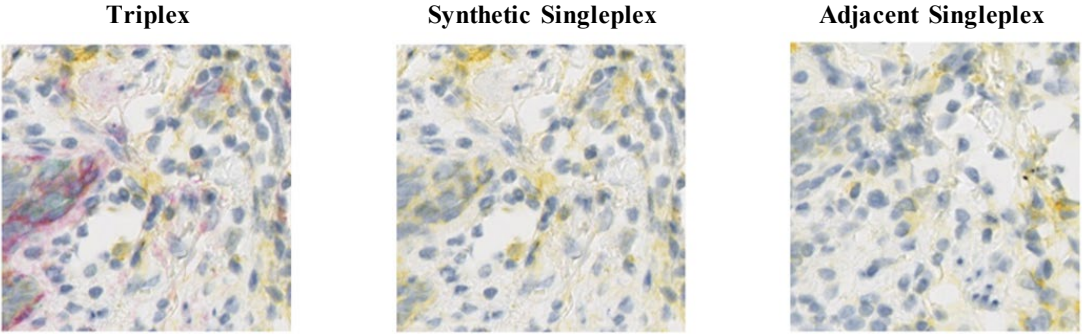

Figure 2: Results on QM-Dabsyl (cMET-PDL1-EGFR, Lung)

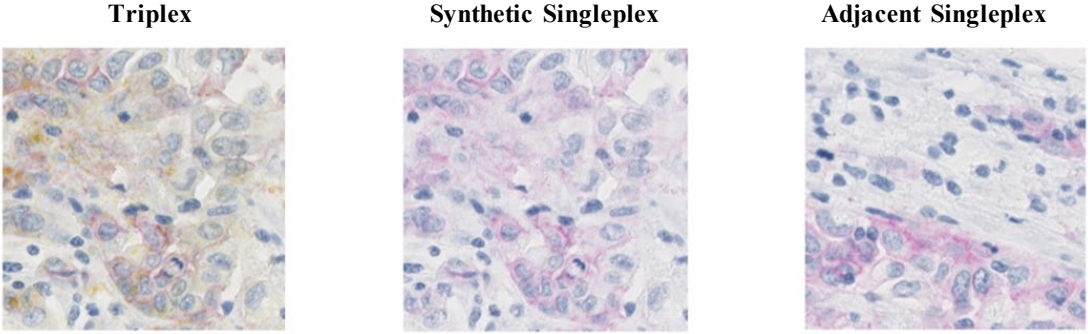

Figure 3: Results on Tamra (cMET-PDL1-EGFR, Gastric)

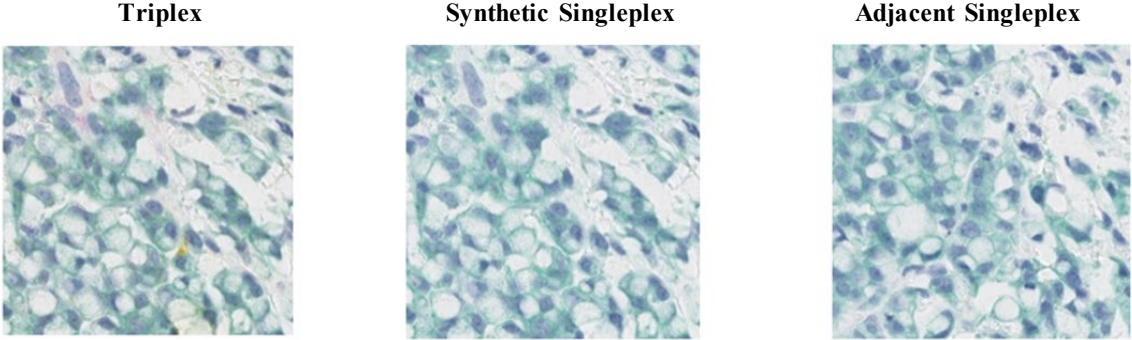

Figure 4: Results on Green (cMET-PDL1-EGFR, Colon)

To ensure the validity of the method quantitatively, we computed stain intensity histogram correlations in the optical density domain for each of the stains obtained from cycle-GAN. We also compared it with an existing non-negative matrix factorization (NMF) method [2] which showed that cycle-GAN performed better than the NMF method and was very competent in creating singleplexes for each of the three stains. Table 1 demonstrates the similarity metrics for the methods for cMET-PDL1-EGFR.

Table 1: Stain Intensity Histogram Correlations for cycle-GAN and NMF on cMET-PDL1-EGFR Images

| Image Similarity Measures | Dabsyl Singleplex | Tamra Singleplex | Green Singleplex |
|---|---|---|---|
| Histogram Correlation (cycle-GAN) | 0.9980 | 0.9997 | 0.9864 |
| Histogram Correlation (NMF) | 0.9435 | 0.9167 | 0.8349 |

Two pathologists also scored the different stain intensity patterns within tumor regions for both the assays. An example of pathologist scoring is shown in Figure 5. We shared adjacent and synthetic singleplex images side by side without letting them know which one was adjacent and which one was synthetic. The figure shows scores from one pathologist for various levels of intensities like 'No Stain', 'Weak' stains and Strong/Moderate' stains within tumor regions of Adjacent and Synthetic Dabsyl singleplexes respectively from the cMET-PDL1-EGFR assay.

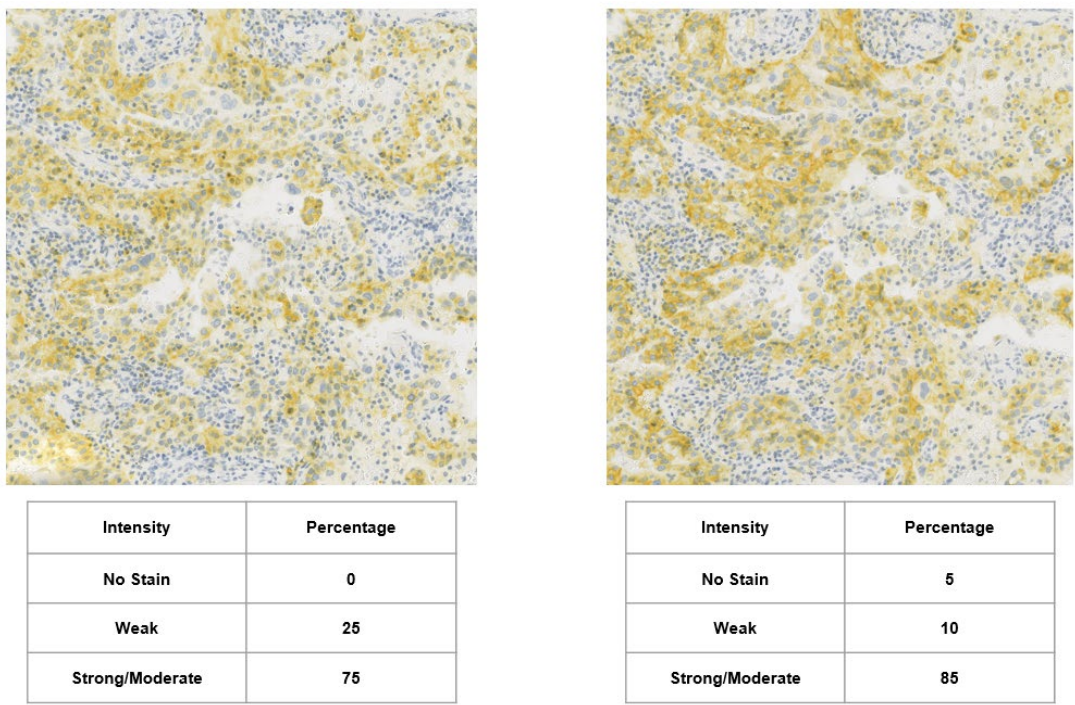

Figure 5: Example of Pathologist Scoring

Upon acquiring scores from two pathologists, we plotted the scores for both the adjacent and synthetic singleplexes superimposed for different levels of intensities of each of the stains. Figures 6 and 7 show the scores for varying stain intensities of Green from two different pathologists for cMET-PDL1-EGFR. The blue curves correspond to the pathologist scores for the adjacent singleplex images whereas the orange curves correspond to that of the synthetic singleplexes. We observed that both the pathologists were pretty much aligned with respect to the scores of the adjacent versus the synthetic singleplexes.

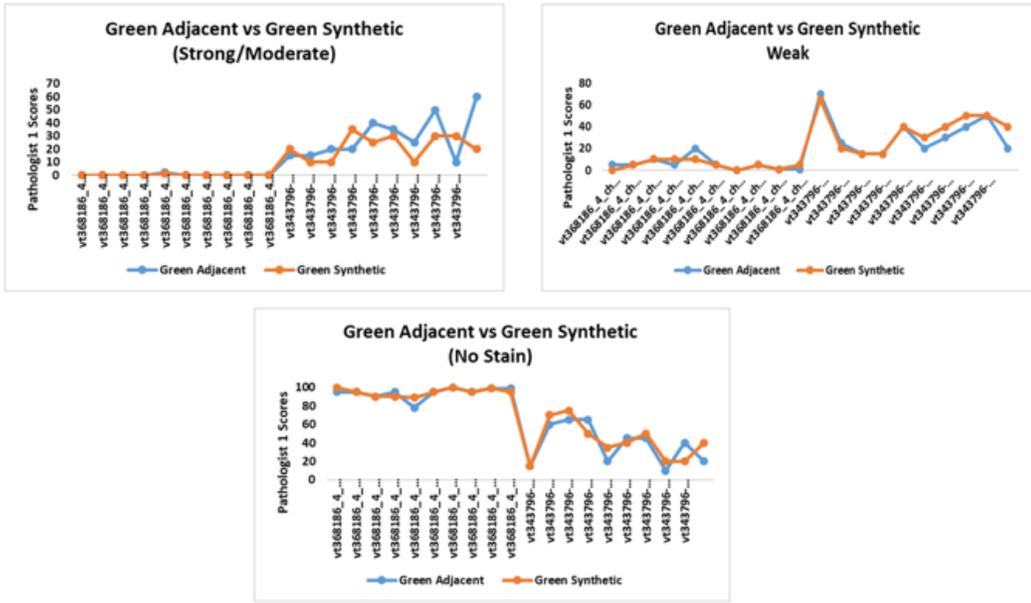

**Figure 6: Plot of Scores from Pathologist 1 for Adjacent and Synthetic Green Singleplexes for Varying Stain Intensities**

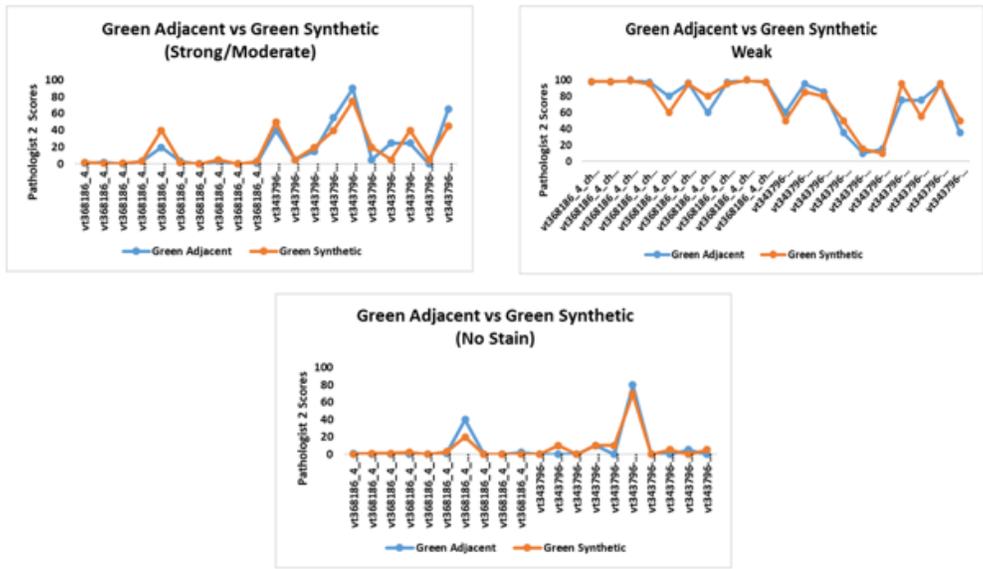

**Figure 7: Plot of Scores from Pathologist 2 for Adjacent and Synthetic Green Singleplexes for Varying Stain Intensities**

Figures 8 and 9 show the scores for varying stain intensities of Dabsyl from two different pathologists for cMET-PDL1-EGFR. Though we could see some differences between scores of two pathologists, the differences were minor and also the scores from each of the pathologists for the adjacent and synthetic singleplexes were pretty much coherent.

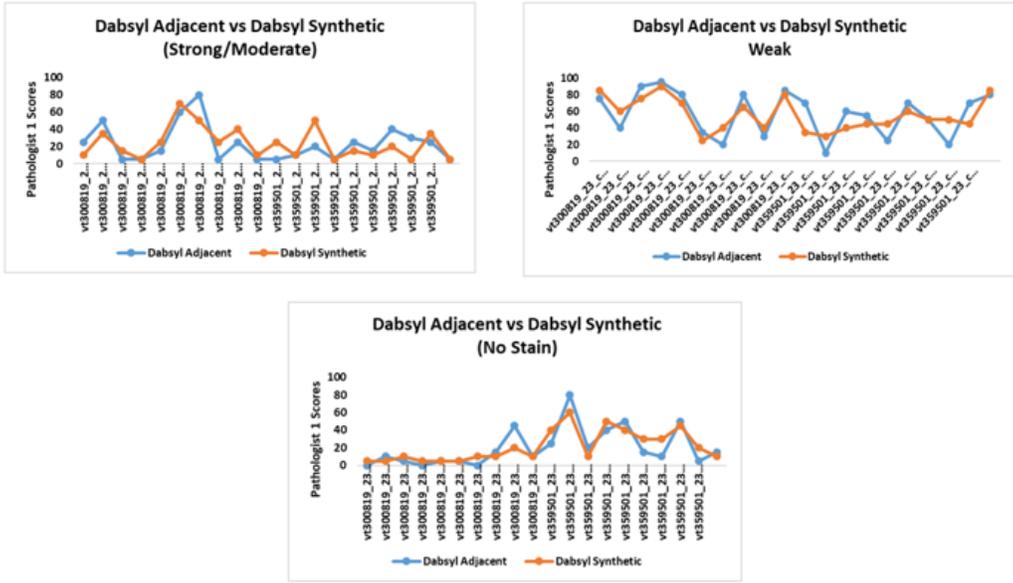

Figure 8: Plot of Scores from Pathologist 1 for Adjacent and Synthetic Dabsyl Singleplexes for Varying Stain Intensities

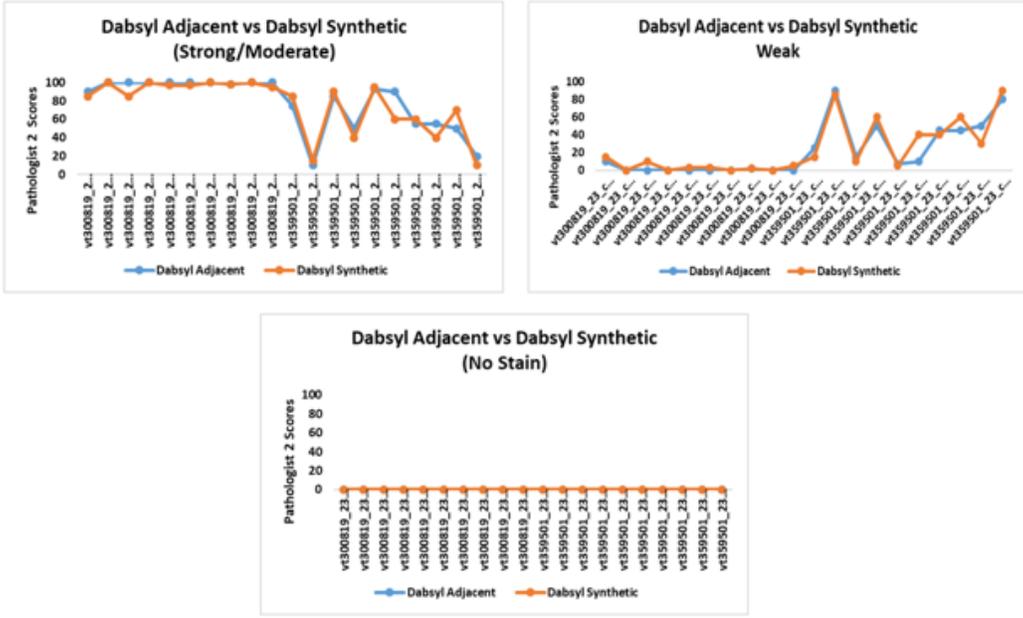

Figure 9: Plot of Scores from Pathologist 2 for Adjacent and Synthetic Dabsyl Singleplexes for Varying Stain Intensities

Figures 10 and 11 show the scores for varying stain intensities of Tamra from two different pathologists for cMET-PDL1-EGFR. We have observed the best results for Tamra. Both the pathologists were very well aligned with respect to the scores of the adjacent versus the synthetic singleplexes as well as among the scores between themselves.

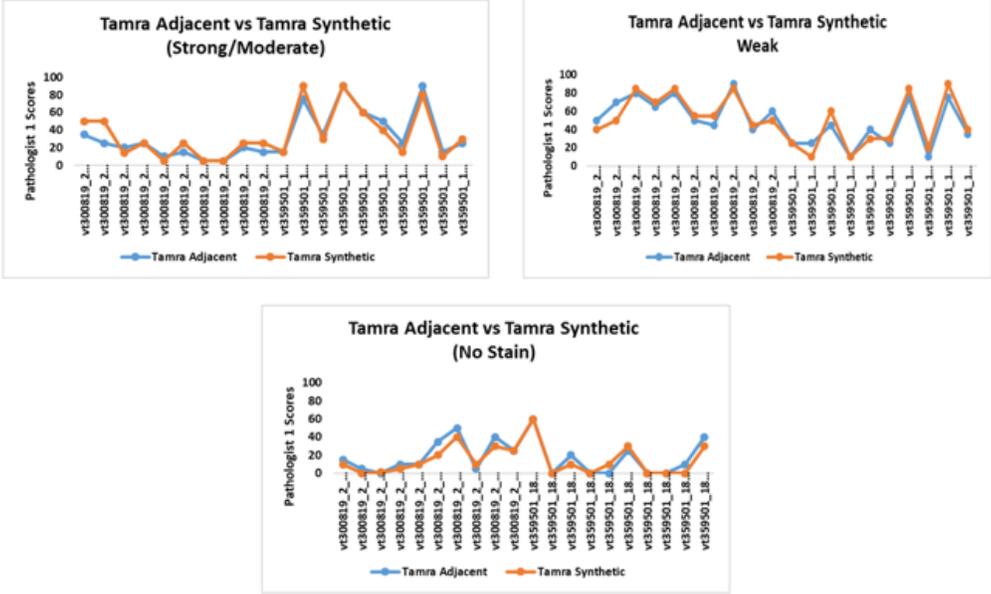

**Figure 10: Plot of Scores from Pathologist 1 for Adjacent and Synthetic Tamra Singleplexes for Varying Stain Intensities**

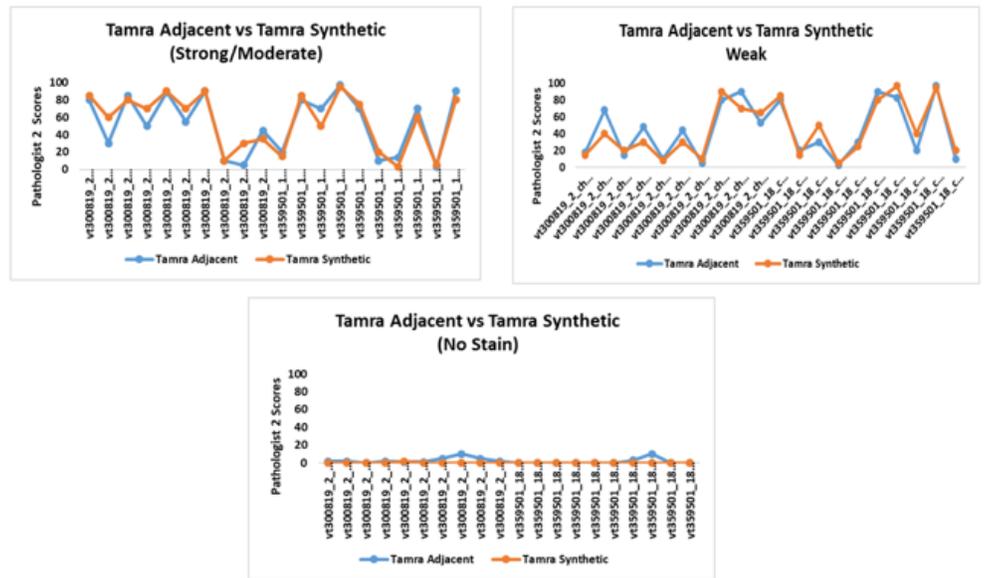

**Figure 11: Plot of Scores from Pathologist 2 for Adjacent and Synthetic Tamra Singleplexes for Varying Stain Intensities**

## 3.2. Results on CD8-LAG3-PDL1

In this section, visual and quantitative results from cycle-GAN as well as pathologist evaluations are presented for the CD8-LAG3-PDL1 assay.

Figures 12-14 show visual results of the method for all the three stains. The leftmost image in each of the figures shows the triplex, the middle image shows the generated synthetic singleplex obtained from the cycle-GAN and the rightmost image shows the corresponding adjacent singleplex image.

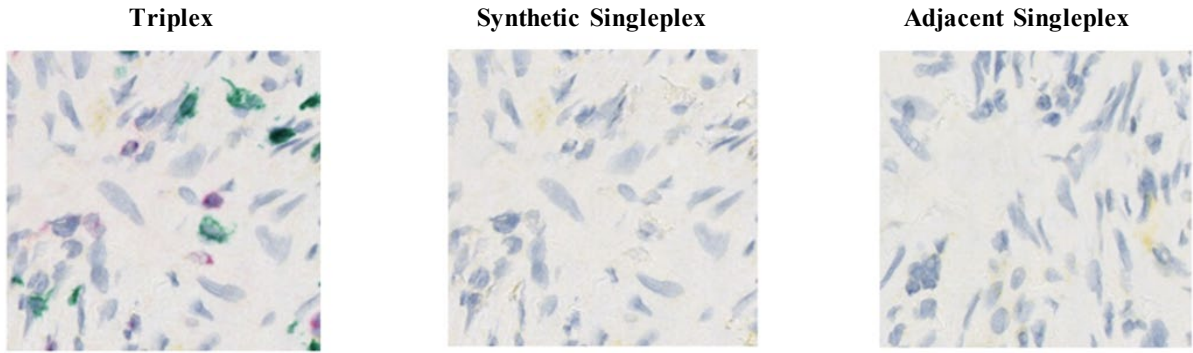

**Figure 12: Results on QM-Dabsyl (CD8-LAG3-PDL1, Colon)**

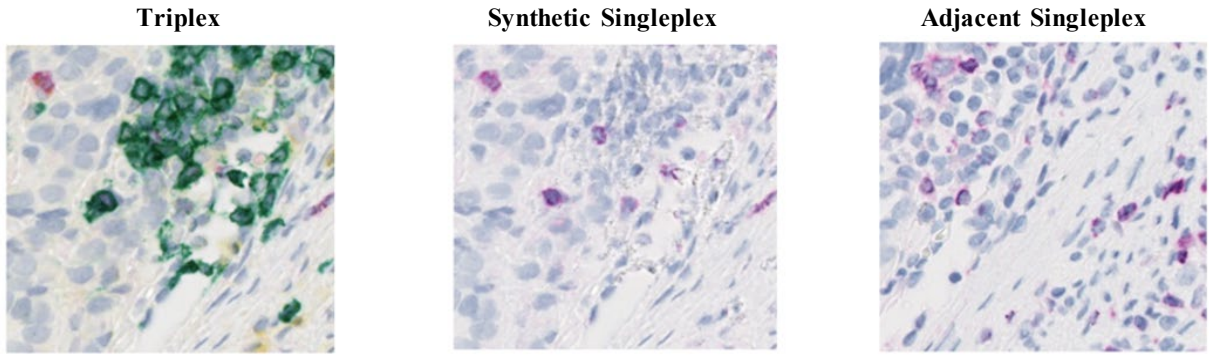

**Figure 13: Results on Tamra (CD8-LAG3-PDL1, Colon)**

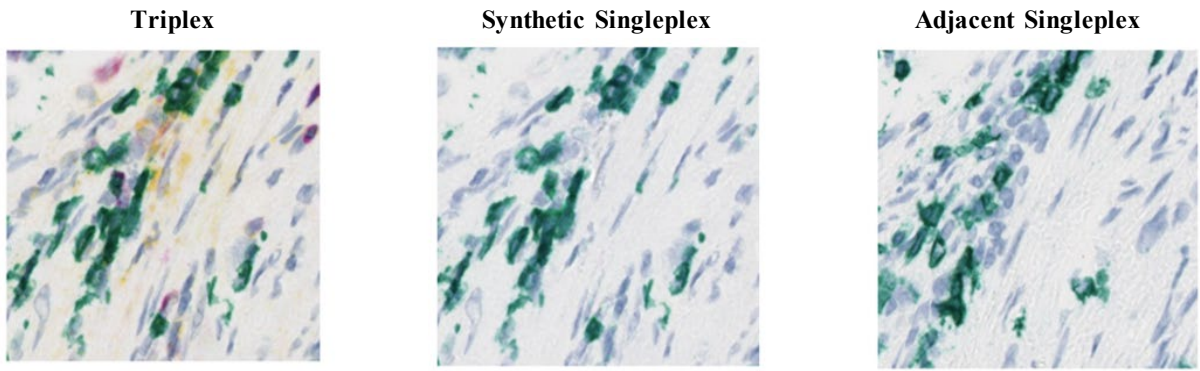

**Figure 14: Results on Green (CD8-LAG3-PDL1, Colon)**

Same as that to the cMET-PDL1-EGFR assay, we computed stain intensity histogram correlations in the optical density domain for each of the stains of the CD8-LAG3-PDL1 assay obtained from cycle-GAN and compared with the NMF method which showed that cycle-GAN performed better than the NMF method and was very competent in creating singleplexes for each of the three stains. Table 2 demonstrates the similarity metrics for the methods for CD8-LAG3-PDL1.

Table 2: Stain Intensity Histogram Correlations for cycle-GAN and NMF on CD8-LAG3-PDL1 Images

| Image Similarity Measures | Dabsyl Singleplex | Tamra Singleplex | Green Singleplex |
|---|---|---|---|
| Histogram Correlation (cycle-GAN) | 0.9837 | 0.9971 | 0.9864 |
| Histogram Correlation (NMF) | 0.9712 | 0.9789 | 0.8056 |

Similar to cMET-PDL1-EGFR, we collected scores from two pathologists for the CD8-LAG3-PDL1 assay and plotted the scores for both the adjacent and synthetic singleplexes overlaying on each other for different levels of intensities of each of the stains. Figures 15 and 16 show the scores for varying stain intensities of Green from two different pathologists for CD8-LAG3-PDL1. The blue curves correspond to the pathologist scores for the adjacent singleplexes whereas the orange curves correspond to that of the synthetic singleplexes. We observed that both the pathologists were pretty much aligned with respect to the scores of the adjacent versus the synthetic singleplexes. They were also quite aligned between themselves regarding the scores of 'Strong/Moderate' stain and 'No stain'. There were some minor disagreements among the scores of the 'Weak' stains which were within acceptable range.

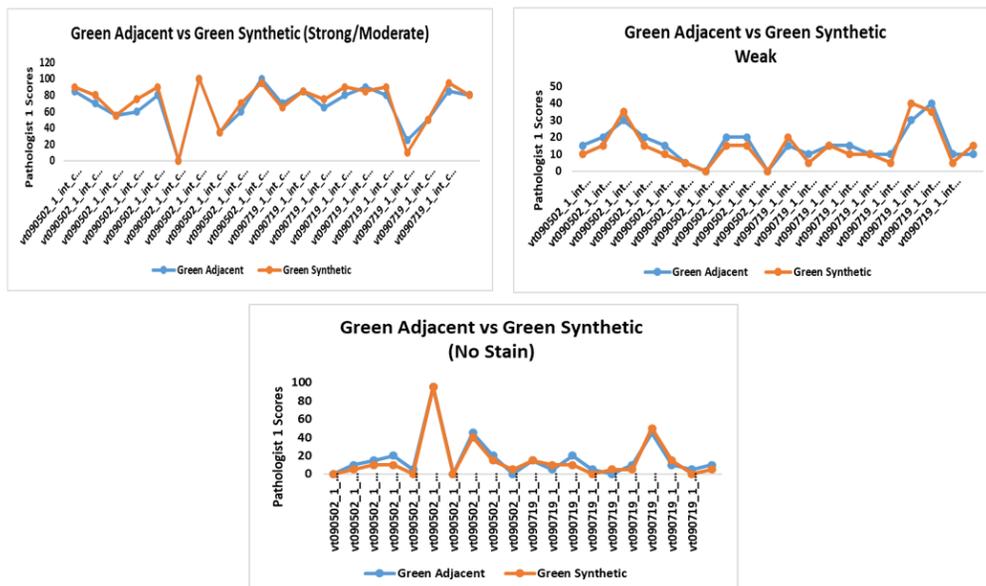

**Figure 15: Plot of Scores from Pathologist 1 for Adjacent and Synthetic Green Singleplexes for Varying Stain Intensities**

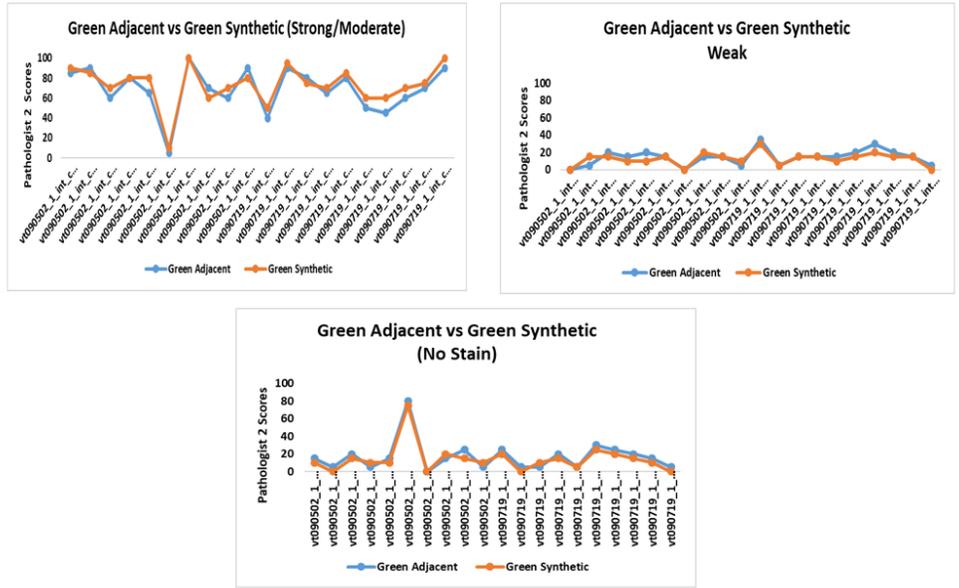

**Figure 16: Plot of Scores from Pathologist 2 for Adjacent and Synthetic Green Singleplexes for Varying Stain Intensities**

Figures 17 and 18 show the scores for varying stain intensities of Dabsyl from two different pathologists for CD8-LAG3-PDL1. We observed some differences between scores of two pathologists from the plots. But these differences were minor and within the acceptable range. Also, the scores from each of the pathologists for the adjacent and synthetic singleplexes were pretty much coherent.

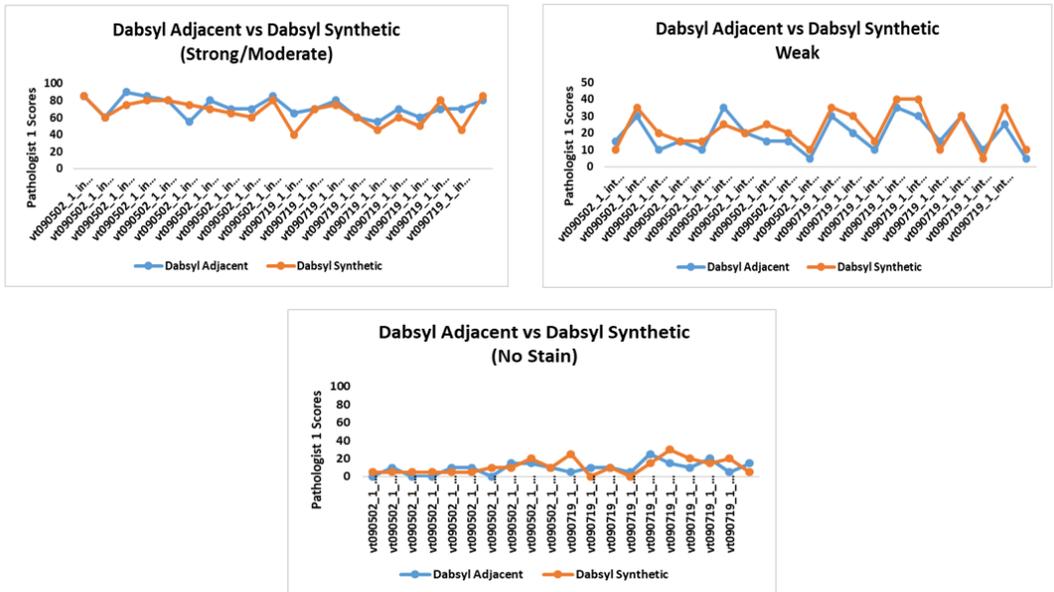

**Figure 17: Plot of Scores from Pathologist 1 for Adjacent and Synthetic Dabsyl Singleplexes for Varying Stain Intensities**

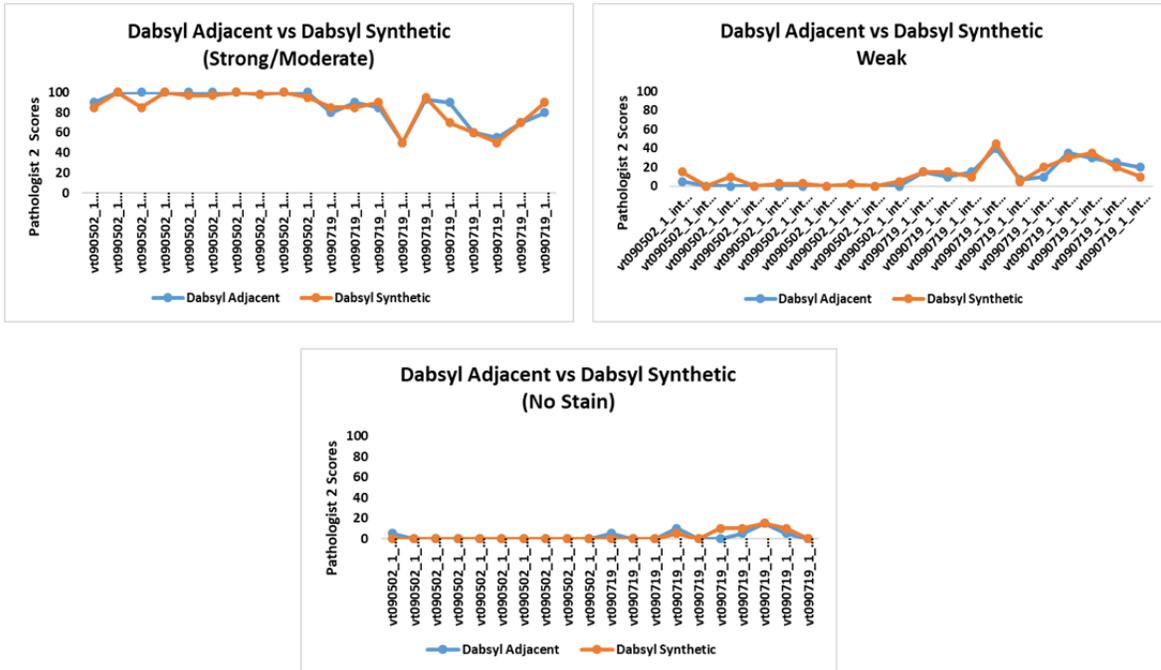

**Figure 18: Plot of Scores from Pathologist 2 for Adjacent and Synthetic Dabsyl Singleplexes for Varying Stain Intensities**

Figures 19 and 20 show the scores for varying stain intensities of Tamra from two different pathologists for CD8-LAG3-PDL1. For Tamra, we observed that both the pathologists were very well aligned with respect to the scores of the adjacent versus the synthetic singleplexes as well as among the scores between themselves.

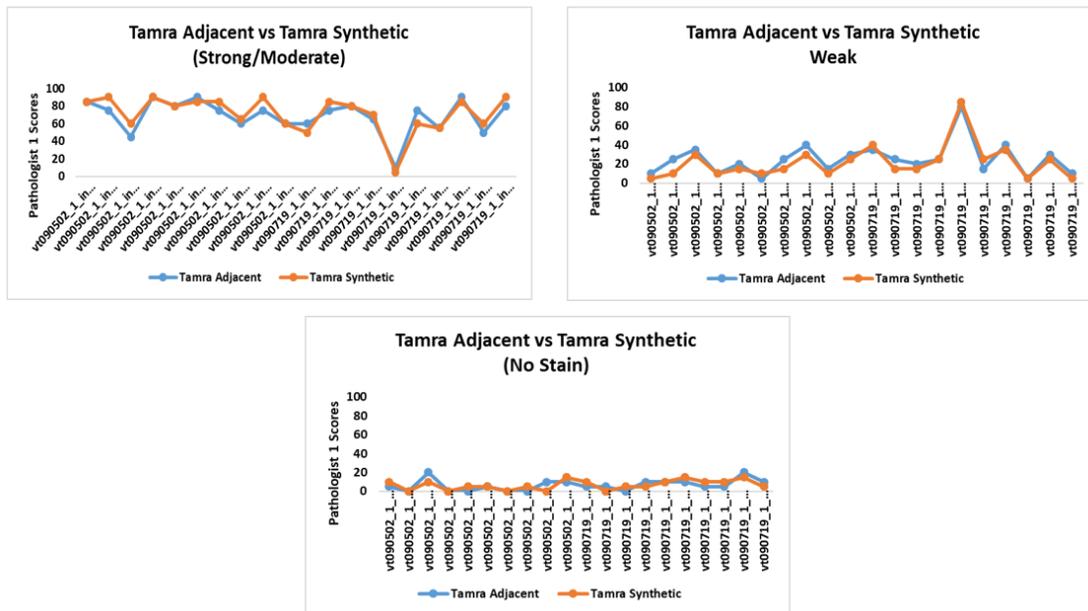

**Figure 19: Plot of Scores from Pathologist 1 for Adjacent and Synthetic Tamra Singleplexes for Varying Stain Intensities**

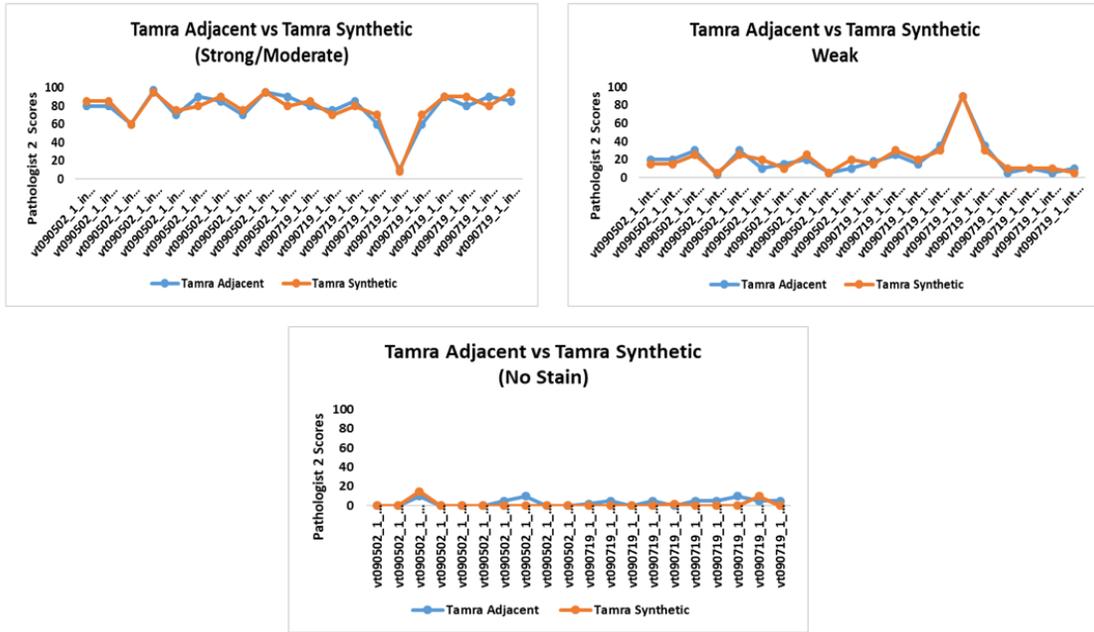

**Figure 20: Plot of Scores from Pathologist 2 for Adjacent and Synthetic Tamra Singleplexes for Varying Stain Intensities**

Last but not the least, in order to quantitatively measure the consensus of scores between the two pathologists, we randomly selected 20 adjacent singleplex FOVs (10 from each of the assay) and computed the median of their scores for the Strong/Moderate stain. Figure 21 shows the plot of the consensus where the blue dots represent the median of the scores i.e. the consensus and the red bars represent the error bars. We observed that in spite of some disagreements (which were within acceptable range), the pathologists were pretty aligned between each other with respect to their scores.

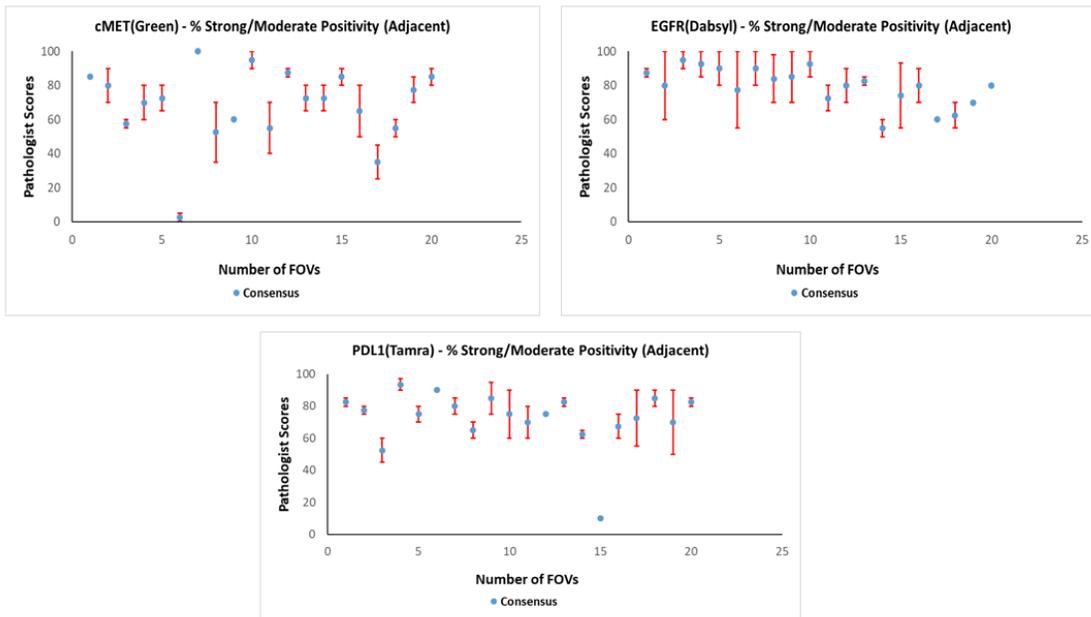

**Figure 21: Plot of Pathologist Consensus for Strong/Moderate Stains of Green, Dabsyl and Tamra Singleplexes**

## 4. CONCLUSIONS

In this study, we developed a cycle-GAN method for color unmixing multiplex brightfield images. The biomarker sets selected for the study were cMET-PDL1-EGFR and CD8-LAG3-PDL1, both of which had co-localization of membranes from all the three biomarkers. Our experiments demonstrated that the synthetic singleplexes generated by the method had strong correlations with the adjacent singleplexes based on visual and quantitative assessments as well as pathologist evaluations. Thus, we can claim that the method is highly efficient, can robustly generate synthetic singleplex images across different contexts and can be readily applied to any other multiplex assay using chromogens of similar color combinations.